\documentclass[12pt,showpacs,showkeys,preprintnumbers,nofootinbib]{article}
\usepackage[english]{babel}
\usepackage[utf8]{inputenc}

\usepackage{amssymb,amsmath,graphics,graphicx}
\title{Effect of undecided agents on an opinion-forming model}
\author{Victor H. Blanco, Ver\'onica Calder\'on\\
\small{Facultad de Ciencias Puras y Naturales, UMSA, La Paz - Bolivia}}

\begin{document}

\maketitle
\begin{abstract}
The effect of undecided agents is studied within populations in an opinion-forming dynamic, varying the number of undecided agents for different proportions of populations in a complete opinion-exchange network. The result is that the dynamic depends on the number of undecided agents, with 10\% of the undecided population potentially affecting the change in consensus and then becoming linear with a negative slope.\\
\textbf{keywords}, Social Systems Dynamics, Collective Phenomenon, Random Processes, Non-Equilibrium Transition Phases, Computational Simulations
\end{abstract}
\section{Introduction}
\begin{figure}[!ht]
\begin{center}
 \includegraphics[scale=0.6]{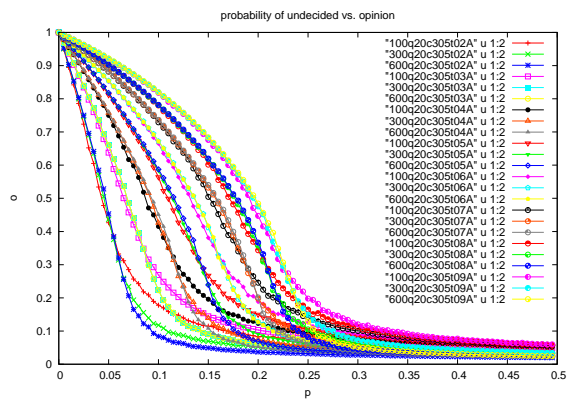}
 \caption{The figure shows agents interacting among themselves obtaining some consensus and observing some fixed points depending on the population of the system, note that the results were scaled.}
 \label{opin}
\end{center}
\end{figure}

\begin{figure}[!ht]
\begin{center}
 \includegraphics[scale=0.6]{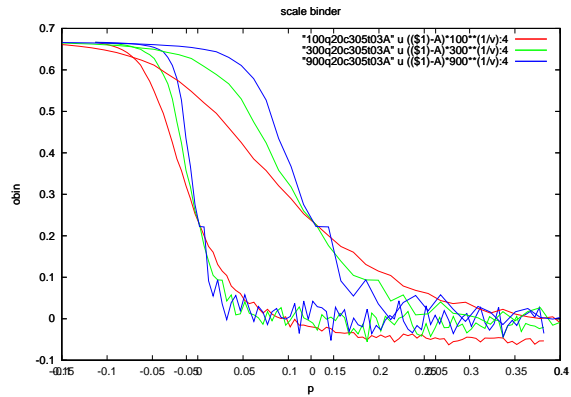}
 \caption{The figure shows the binder for the system, note that the results were scaled.}
 \label{binder}
\end{center}
\end{figure}

The dynamics of opinion on is studied with different models, several shown in \cite{castellanos, Fortunato, Catanzaro}, with different dynamics of opinion formation.
From the results obtained in \cite{nuno} in the model used, and showing the scale behavior in the figures: (\ref{opin}) for the average opinion, (\ref{vari}) its variance and (\ref{binder}) third moment, where is.
he dynamics of the project have been scaled up, and its exponents emerge: caracterısticos. It has been observed that in different situations prior to an election, a certain number of agents have a choice of the moment, their decision not depending on a memory from which they extract their final decision, but if these agents during the course can influence other agents during the dynamics \cite{Abrahamsson,Ghaderi}. Now I wish to see the effect of undecided agents who, without a preference for choice, develop their choice without the influence of other agents, but whose decision is totally random, nevertheless, on their way to choosing their preference of the moment.
To observe what is effect on a decision of the group. Based on three possible choices, if they influence others but being undecided they change their influence in their moment. also the variance and the binder: With their respective scale exponents for each magnitude. Now it is tried to use the model in a situation already agreed to observe the effect of the undecided agents in that decision.
the variance and the binder too:
\begin{figure}[!ht]
\begin{center}
 \includegraphics[scale=0.6]{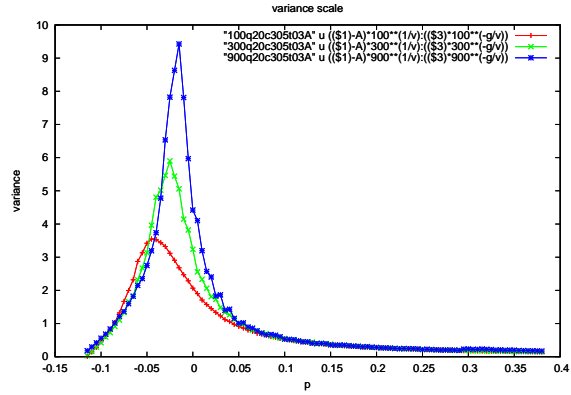}
 \caption{The figure shows the variance of the agents interacting in the opinion system, note that the results were scaled}
 \label{vari}
\end{center}
\end{figure}
With their respective scale exponents for each magnitude. Now it is tried to use the model in a situation already agreed to observe the effect of the undecided agents in that decision.
With their respective scale exponents for each magnitude.
\section{Model}
\begin{figure}[!ht]
\begin{center}
 \includegraphics[scale=0.7]{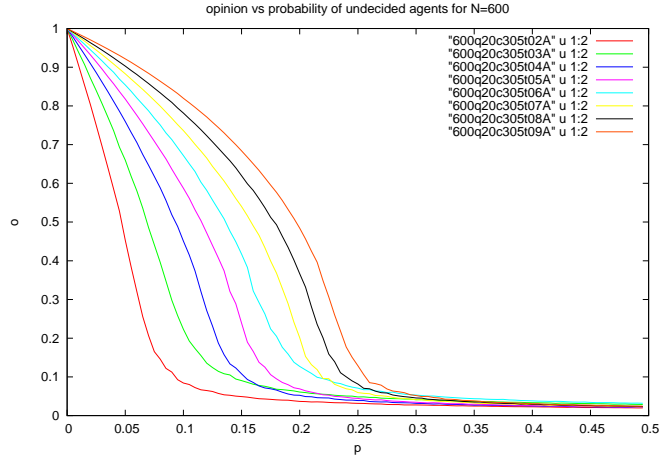}
 \caption{The figure shows the opinion of the group varying the number of undecided agents of probability from $0.2$ to $0.9$}
 \label{fig:opin600i}
\end{center}
\end{figure}

Similar to the model used in \cite{nuno}, a fraction of the population is now associated as volatile in different proportions to the dynamics used, with the following considerations:
\begin{itemize}
\item A pair of $i,j$ agents are randomly selected.
\item If $i$ is a volatile agent, it will choose its next opinion randomly from the two possible options.
\item On the other hand if it is a non-volunteer agent it will interact with its opinion for the time $t+1$ with the following rule:
\begin{displaymath}
o_i(t+1)=[o_i(t)+\mu_{ij}o_j(t)]
\end{displaymath}
\end{itemize}
$i,j=1,...,N$. For $o_i$ normal individuals. Undecided agents are given by the function::
\begin{equation}
o_i(t+1)=rnd[-1,1] 
\end{equation}
With $rnd$ random function between the two options for undecided agents, that no matter who they interact with, their answer is always random.\\
The proportion of undecided agents varies and they are allowed to interact for different amounts of populations.
\begin{equation}
O=\left<\frac{1}{N}\sum^N_{i=1}o_i\right>
\end{equation} 
Where $<.>$ denotes an average configuration disorder, it is sensitive to the balance condition between extreme views. Note that $O$ plays the role of 'magnetic spin'.
Results were obtained for different populations. The following figure shows how the consensus varies with the introduction of undecided agents in the figure (\ref{fig:opin600i}) taking a population of $600$ agents and varying the probability that these agents are undecided agents. As the case has aleatorical variables, It is not considered more than the average for the analysis.
It can be observed that the fewer undecided agents there are, the more consensus there is in the group and the more group decisions decrease, the greater the probability that there are more undecided agents.\\
Here a case is analyzed, the case of continuous opinions, since it was demonstrated that the different cases of discreet opinions, discreet and continuous, are obtained similar tendency (\cite{nuno}), where it is observed that the dynamic differences with different population in characteristic points are crossed.
\begin{figure}[!ht]
\begin{center}
 \includegraphics[scale=0.7]{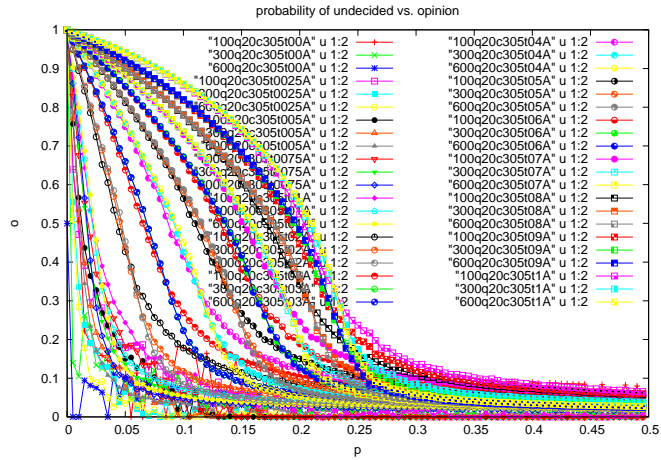}
 \caption{The figure shows how the consensus is for different numbers of undecided agents converging to a point independent of population size.}
 \label{fig:consensoTOT}
\end{center}
\end{figure}
Now introducing other populations the figure (\ref{fig:consensus}) is obtained where it is observed that there is a point of inflection through which the different populations pass, this point will delimit what is called disorder of the election against the order of the same one and the different probabilities of obtaining undecided agents are varied.\\
Elaborating a regression to the system obtained from the transitions from order to disorder, is obtained:
\begin{figure}[!ht]
\begin{center}
 \includegraphics[scale=0.5]{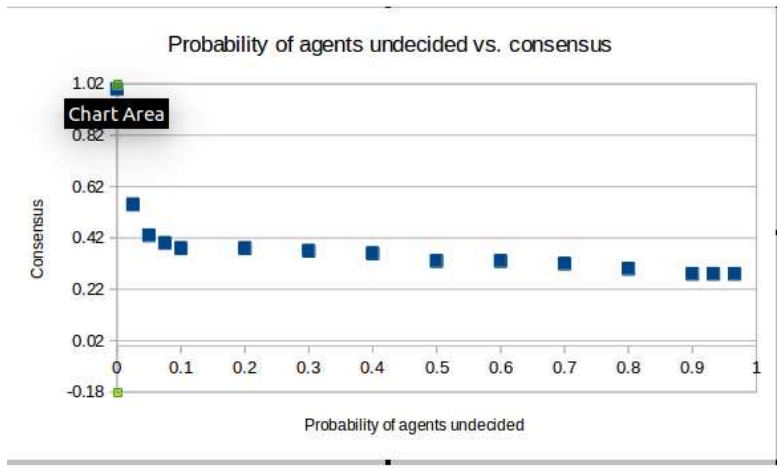}
 \caption{The figure shows how consensus for different numbers of undecided agents is the probability of having undecided agents vs. consensus.}
 \label{fig:consensus}
\end{center}
\end{figure}
De la figura (\ref{fig:consensus}) se obtiene dos tendencias divididas por la probabilidad de agentes indecisos del $0.1$, donde se obtiene:
\begin{figure}[!ht]
\begin{center}
 \includegraphics[scale=0.5]{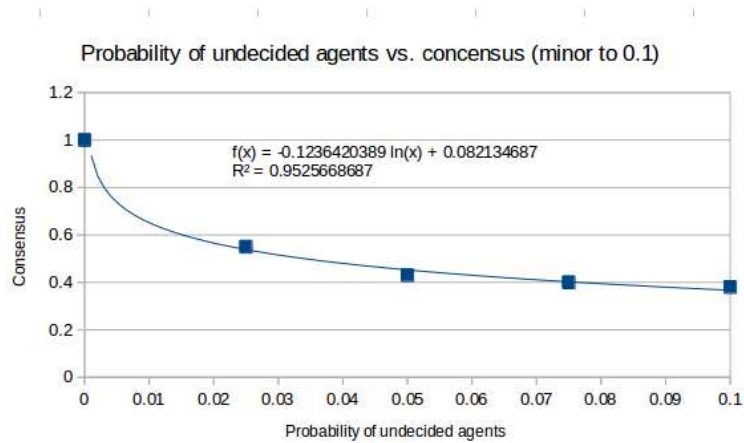}
 \caption{Tendencia potencial de la caida del concenso entre 0 a 0.1 de probabilidad de tener agentes indecisos.}
 \label{fig:menor01}
\end{center}
\end{figure}
From the figure (\ref{fig:consensus}) two trends are obtained divided by the probability of undecided agents of $0.1$, where it is obtained:
\begin{equation}\label{A}
f(x)=0.2x^{0.268} \quad\text{for } x \in [0,0.1]
\end{equation}
Where $f(x)$ represents the consensus reached and $x$ the different proportions of undecided agents that form a relationship for the region between $0$ to $0.1$ probability of undecided agents with potential trend.
\begin{figure}[!ht]
\begin{center}
 \includegraphics[scale=0.5]{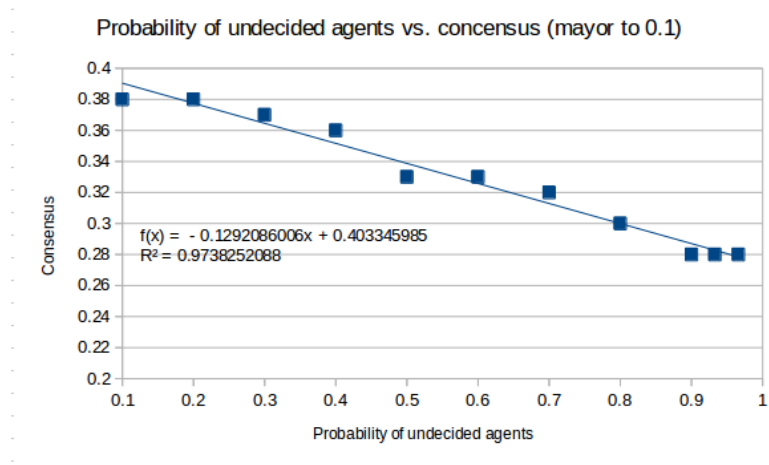}
 \caption{Potential trend of falling consensus between $0$ and $0.1$ probability of having undecided agents. Note that for about $0.035$ of the undecided population, the consensus is affected to half of the total population.}
 \label{fig:mayor01}
\end{center}
\end{figure}
For probability values greater than $0.1$, the consensus behaves linearly with the trend:
\begin{equation}\label{B}
f(x)=-0.129x+0.403 \quad\text{for } x \in [0.1,1]
\end{equation}
For the region greater than $0.1$ the trend tends to be linear, a relationship that obtains a correlation coefficient of $R^2=0.96$, this image shows the border between order with disorder caused by undecided agents. It can be seen that undecided agents cause a greater effect on the consensus of the population when they are less than 10\% of the population. This consensus reaches 30\% of the population when practically the entire population is undecided, an extreme case that would appear to be a society with information coming only from its interactions.
\section{Conclusions}
The influence of volatile decision agents, shows that the 10\% of the undecided agents cause more influence in the consensus that while more the undecided agents increase. For a few undecided agents, the effect is greater, obeying a potential fall $f(x)=0.2x^{0.268}$ causing decisions to move away from unanimous consensus. For higher values of undecided agents, the trend of this relationship behaves with a slope: $-0.129$ continues.\\
Undecided agents can cause a change in the trend of the consensus as they increase, for this 3.5\% of undecided agents of the total population is enough to decrease the initial consensus by half, which represents as a practice that causing doubts in the population can cause a turn of the trend of the consensus.
\section*{Acknowledgments}
A very special thank you to Celia Anteneodo for her infinite patience.\\

\end{document}